\documentclass[english,aps,prb,superscriptaddress]{revtex4-1}
\usepackage{babel}
\usepackage{amsmath,mathtools}
\usepackage{amssymb}
\usepackage{graphicx}
\begin{document}
\title{Non-Hermitian topological phases and exceptional lines in topolectrical circuits}
\author{S M Rafi-Ul-Islam }
\email{e0021595@u.nus.edu}
\selectlanguage{english}%
\affiliation{Department of Electrical and Computer Engineering, National University of Singapore, Singapore}
\author{Zhuo Bin Siu}
\email{elesiuz@nus.edu.sg}
\selectlanguage{english}%
\affiliation{Department of Electrical and Computer Engineering, National University of Singapore, Singapore}
\author{Mansoor B.A. Jalil}
\email{elembaj@nus.edu.sg}
\selectlanguage{english}%
\affiliation{Department of Electrical and Computer Engineering, National University of Singapore, Singapore}
\begin{abstract}
We propose a scheme to realize various non-Hermitian topological phases in a topolectrical (TE) circuit network consisting of resistors, inductors, and capacitors. These phases are characterized by topologically protected exceptional points and lines. The positive and negative resistive couplings $R_g$ in the circuit provide loss and gain factors which break the Hermiticity of the circuit Laplacian. By controlling $R_g$, the exceptional lines of the circuit can be modulated, e.g., from open curves to closed ellipses in the Brillouin zone. In practice, the topology of the exceptional lines can be detected by the impedance spectra of the circuit. We also considered finite TE systems with open boundary conditions, the admittance spectrum of which exhibits highly tunable zero-admittance states demarcated by boundary points (BPs). The phase diagram of the system shows topological phases which are characterized by the number of their BPs. The transition between different phases can be controlled by varying the circuit parameters and tracked via impedance readout between the terminal nodes.  Our TE model offers an accessible and tunable means of realizing different topological phases in a non-Hermitian framework, and characterizing them based on their boundary point and exceptional line configurations.
\end{abstract}
\maketitle
\subsection{Introduction}
There is growing interest in studying topological states in various platforms such as topological insulators\cite{ref_1,ref_2}, cold atoms\cite{ref_3,ref_4}, photonics systems \cite{ref_5,ref_6}, superconductors\cite{ref_7}, and optical lattices\cite{ref_8,ref_9} due to their extraordinary properties such as topologically protected edge states and unconventional transport characteristics \cite{rafi2020anti,rafi2020strain}. Such topological states have been studied in Hermitian and lossless systems\cite{ref_10}, where the eigenenergies are always real. However, Hermitian systems do not exhibit many interesting phenomena such as exceptional points\cite{ref_11}, the skin effect\cite{ref_12,ref_13,rafi2021topological}, biorthogonal bulk polarization \cite{ref_14}, and wave amplification and attenuation \cite{ref_15,ref_16}. In the pursuit of more exotic characteristics in topological phases, researchers have shifted attention from Hermitian to non-Hermitian systems\cite{ref_17}. In contrast to Hermitian systems, non-Hermitian systems in general exhibit complex eigenvalues unless the system obeys some specific symmetries such as the $\mathcal{PT}$ symmetry where $\mathit{P}$ and $\mathcal{T}$ are the parity and time reversal operations, respectively. One iconic feature of non-Hermitian systems is the existence of exceptional points, where two or more eigenvectors coalesce and the Hamiltonian becomes nondiagonalizable. This feature leads to  many novel transport phenomena, such as unidirectional transparency\cite{ref_18,ref_19}, unconventional reflectivity \cite{ref_20}, and super sensitivity\cite{ref_21}. 
Generally, the exchange of energy or particles between lattice points and the surrounding environment induces non-Hermiticity in the system.  One way to induce non-Hermiticity is to add imaginary onsite potentials at different sublattices that represent gain or loss in the system depending on the sign of the potentials. In addition, asymmetric sublattice couplings may also induce non-Hermiticity in the system Hamiltonian \cite{rafi2021topological}. However, realizing non-Hermitian systems in condensed matter\cite{ref_22}, acoustic metamaterials\cite{ref_23}, and optical structures\cite{ref_24,ref_25} is in practice difficult because of the limited control over sublattice couplings, instability of the complex eigenspectra\cite{ref_26}, and limitations in experimental accessibility.  
In pursuit of alternative platforms to overcome the aforementioned experimental limitations, topolectrical  (TE) circuits \cite{rafi2020realization,ref_28,ref_29,ref_30,rafi2020topoelectrical} have emerged as an ideal platform to not only realize non-Hermitian systems, but also to investigate many emerging phenomena such as Chern insulators\cite{ref_31}, the quantum spin Hall effect\cite{ref_32,sun2020spin,ref_33,sun2019field}, higher-order topological insulators\cite{ref_34,ref_35}, topological corner modes\cite{ref_29,ref_36}, Klein tunneling \cite{rafi2020anti,rafi2020topoelectrical} and perfect reflection \cite{rafi2020anti,rafi2020strain}. Appropriately designed TE circuits can emulate the topological properties of materials and offer unparalleled degrees of tunability and experimental flexibility through the conceptual shift from conventional materials system to artificial electrical networks. The freedom in design and control over lattice couplings allow us to investigate electronic structures beyond the limitations of condensed matter systems. Moreover, TE circuit networks are not constrained by the physical dimension or distance between two lattice (nodal) points but are described solely by the mutual connectivity of the circuit nodes. Besides, the freedom of choice in the connections at each node and long-range hopping make TE systems easier to fabricate compared to real material systems.  Therefore, it is also possible to design an equivalent circuit network in lower dimensions that resembles the characteristics of higher-dimensional circuits \cite{rafi2020topoelectrical}. Unlike real material systems, an infinite real system can be mimicked by finite-sized circuit networks in TE circuits. Therefore, TE networks enable us to design a non-Hemitian system in a $\mathrm{RLC}$ circuit network with better measurement accessibility as all the characteristic variables such as the admittance bandstructure and the density of states can be evaluated through electrical measurements\cite{ref_37}(e.g. impedance, voltage, and current readings).
In this paper, we investigate the exceptional lines, i.e. loci of exceptional points (EP), in a non-Hermitian TE system consisting of electrical components such as resistors, inductors and capacitors as a function of the non-Hermitian parameter (i.e. resistance). We show that by introducing non-Hermiticity to the circuit Laplacian through the insertion of positive or negative resistances between the voltage nodes and the ground to induce imaginary onsite potentials in the lattice sites and tuning the non-Hermitian parameters appropriately through using by using appropriate resistance values, the loci of the EPs may be easily switched to take the form of either a line or closed curves such as ellipses in the Brillouin zone. We show that a unique property of the TE platform over earlier works \cite{rui2019pt,choi2019addendum,stegmaier2020topological}, viz. the impedance spectrum,  is that the  exceptional lines can be detected and easily tracked by measuring the impedance spectrum of the circuit.  We further investigate finite systems with open boundary conditions along one direction, and find that these finite systems possess a rich phase diagram with different phases possessing up to four pairs of BPs, depending on the circuit parameters. We also show that edge states in Hermitian or pure $\mathrm{LC}$ circuits, become hybridized with the bulk modes in non-Hermitian RLC circuits. In summary, our TE model provides an experimentally accessible means to investigate the phase diagram and various topological phases of non-Hermitian systems.
\subsection{Theoretical Model}
\begin{figure}[h!]
  \centering
    \includegraphics[width=0.65\textwidth]{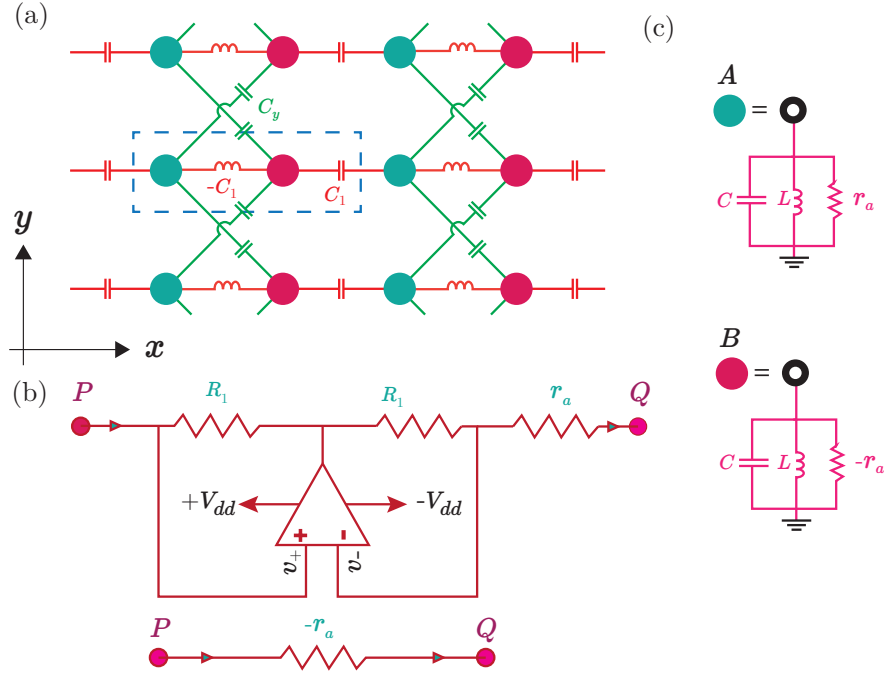}
  \caption{a) Schematic of the two-dimensional TE lattice in the $x$-$y$ plane. The blue and magenta circles represent sublattices $A$ and $B$ respectively.  The alternating sublattice sites $A$ and $B$ are connected to each other in the $x$-direction by an inductor $-C_1$ and capacitor $C_1$  for the intracell and intercell connections, respectively (the dashed rectangle delineates a unit cell). Along the $y$-direction, nodes of different sublattices are connected diagonally by a capacitor $C_y$. b) Schematic circuit of a negative resistance converter, which introduces a $\pi$-phase difference and therefore converts a loss resistive term $r_a$ to a gain term $-r_a$. The combination of two resistors having the same resistance $R_1$ along with an ideal operational amplifier with supply voltages $+V_{dd}$ and $-V_{dd}$ results in current inversion and hence acts as a negative resistance converter. c) Grounding mechanism of the TE circuit. All nodes are connected to ground via a parallel combination of a common capacitor ($C$) and inductor ($L$). Furthermore, in parallel to these, a positive resistor $r_a$ (loss term) and negative resistor $-r_a$ (gain term) is connected to ground from the $A$ and $B$ nodes, respectively. The negative resistor $-r_a$ is implemented by means of INRC depicted in (b).  }
  \label{figsch6}
\end{figure}

We consider a two-dimensional TE circuit, shown in Fig. \ref{figsch6}, which is composed of capacitors, inductors, positive resistors (loss elements) and negative resistors (gain elements). The circuit has a unit cell consisting of two sublattice nodes $A$ and $B$. Each node is connected by a capacitor of $\pm C_1$ and $C_y$ to its nearest neighbour in the $x$ and $y$-direction, respectively, and a parallel combination of a common capacitor $C$ and inductor $L$ to the ground. The inductance $L$ can be varied to modulate the resonance condition. To explore the effects of loss and gain in the TE network, we consider an imaginary onsite potential $i R_g$ on the $A$ nodes and $-i R_g$ on the $B$ nodes. This onsite potential can be realized by connecting resistors of resistance $r_a$ ( $-r_a$) between each $A$ ($B$) node and the ground. The onsite potential are related to the resistors by $R_g=1/{\omega r_a}$, where $\omega$ denotes the frequency of the driving alternating current. The negative imaginary onsite potential at the $B$-type nodes can be obtained by using op-amp-based negative resistance converters with current inversion (INRC) (see Supplemental section 1 for details) Alternatively,  similar gain and loss terms can obtained in our TE model via using only two unequal positive resistances connected to ground for each type of nodes instead of using negative resistance converters, which yields the mathematically similar Hamiltonain as Eq. \ref{WSMham} except for  a global shift of the imaginary part of all the admittance eigenenergies (see Supplementary Note 1 for details). The advantage of using grounding resistors with different values of (positive) resistance is the dynamic stability of the circuit because the absence of op-amps avoids the possibility that the voltage profiles may get over-amplified by the negative resistance.    
 The Laplacian of the TE circuit over the $x$-$y$ plane can be expressed as
\begin{eqnarray}
	H_{2D}({k_x},{k_y}) = - ( C_1(1+\cos{k_x} )+ 2C_y\cos{k_y} )\sigma_x  - C_1\sin{k_x}\sigma_y  + i R_g \sigma_z  ,
 	\label{WSMham}   	
\end{eqnarray}
where the $\sigma_i$s denote the Pauli matrices in the sublattice space. In the absence of resistances, Eq. \ref{WSMham} exhibits  both the chiral symmetry, i.e., $\mathcal{C} H_{2D}(k_x,k_y)\mathcal{C}^{-1}=-H_{2D}(k_x,k_y)$ and inversion symmetry, i.e.,  $\mathcal{I}_n H_{2D}(k_x,k_y)\mathcal{I}_n^{-1}=H_{2D}(-k_x,-k_y)$, where the chirality and inversion operators are $\mathcal{C}=\sigma_z$ and $\mathcal{I}_n=\sigma_x$, respectively. However, a finite  $R_g$ would break the inversion symmetry of the Laplacian in Eq. \ref{WSMham} although the parity time ($\mathcal{PT}$) symmetry as defined by  $\mathcal{P} \mathcal{T} H_{2D}(k_x,k_y)\mathcal{T}^{-1}\mathcal{P}^{-1}=H_{2D}^{*}(k_x,k_y)$, would still be preserved (here, $\mathcal{P}=\sigma_x$ is the parity operator and $\mathcal{T}$ is complex conjugation). Therefore, the Laplacian in Eq. \ref{WSMham} can still have real eigenvalues despite its non-Hermiticity \cite{ref_b}. 
More specifically, a non-zero $R_g$ in Eq. \ref{WSMham} corresponds to the insertion of alternating $i R_g$ and $-iR_g$ terms on the diagonal of the Laplacian, which preserves the commutation with the $\mathcal{PT}$ operator \cite{ref_c}. In contrast to the Hermitian case, $\mathcal{PT}$ symmetry in the non-Hermitian TE system eigenmodes can be broken depending on the model parameters, i.e., the eigenmodes of Eq. \ref{WSMham} are not necessarily the eigenstates of the $\mathcal{PT}$ operator \cite{ref_d} even when the Laplacian itself respects $\mathcal{PT}$ symmetry. In this situation, complex admittance spectra emerge with exceptional points where both the hole- and particle-like admittance bands coalesce. 
Therefore, the complex admittance dispersion for the circuit model take the form of
\begin{eqnarray}
    E_{2D}(k_x, k_y)&=& \sqrt{2C_1^2 (1+\cos{k_x})+ 4 C_y^2 \cos^2 {k_y}+ 4 C_1 C_y \cos{k_y} (1+\cos{k_x}) - R_g^2},
\label{eqe2dch6}  
\end{eqnarray}
where the $\pm$ refers to the two admittance bands, respectively. By tuning the circuit parameters, we can obtain different numbers of real solutions for $\vec{k}$ for $E_{2D}=0$ in Eq. (\ref{eqe2dch6}), which translates into different number of exceptional points in the Brillouin zone (BZ). The exceptional points occur at $\boldsymbol w_{ex}=(k_x,k_y)=(\pi,\pm\arccos( R_g/2C_y))$ and $\boldsymbol w_{ex}=(\pi,\mp\pi\pm\arccos(R_g/2C_y))$. 
\begin{figure}[ht!]
\centering
\includegraphics[scale=0.42]{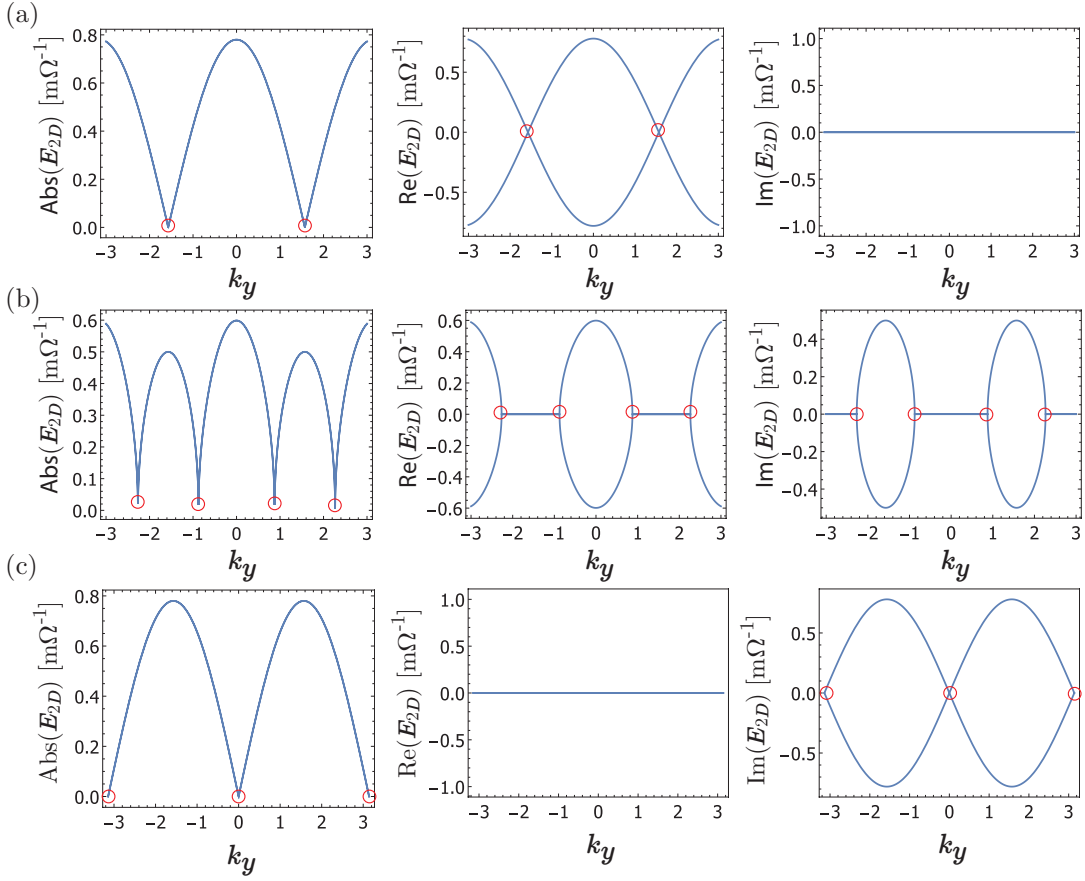}
\caption{ Absolute value, real part, and imaginary part of the complex admittance as a function of $k_y$ with the parameters $C_1=0.78$ $\mathrm{mF}$, $C_y=0.39$ $\mathrm{mF}$, and $k_x=\pi$. We consider three representative values of the grounding resistance, i.e., (a) $R_{g}=0$ $\mathrm{m\Omega^{-1}{Hz}^{-1}}$,  (b) $R_{g}=0.5$ $\mathrm{m\Omega^{-1}{Hz}^{-1}}$, and (c) $R_{g}=0.78$ $\mathrm{m\Omega^{-1}{Hz}^{-1}}$. Note that case (c) corresponds to the critical value of the non-Hermitian parameter $R_{g}=2C_y=0.78$ $\mathrm{m\Omega^{-1}{Hz}^{-1}}$, beyond which the admittance spectrum becomes purely imaginary. All the exceptional points are represented by open red circles. }
\label{gFig1}
\end{figure}
To illustrate the effect of non-Hermitian gain or loss, we plot the admittance spectrum as a function of wavevector $k_y$ and fix $k_x=\pi$ for three representative values of $R_g$. In the absence of gain or loss (i.e., $R_g=0$), the admittance spectrum becomes purely real with two Dirac points or exceptional points (see Fig. \ref{gFig1}a). Because the Laplacian in Eq. \ref{WSMham} obeys chiral symmetry, for a given $k_y$, the admittance eigenvalues always come in pairs with equal magnitude but opposite signs. For non-zero values of $R_g$, the admittance dispersion becomes complex and the band-touching degeneracy points split into pairs of band-touching exceptional points. For instance, in Fig. \ref{gFig1}b, the splitting of the degeneracy points is most evident in the admittance plots in the left-most and middle columns. Here, the degeneracy point with zero admittance at $k_y \approx -1.6$ in  Fig. \ref{gFig1}a splits into two exceptional points (EPs) at $k_y \approx -2.3$ and $k_y \approx -0.9$ in Fig. \ref{gFig1}b. The admittance spectrum is then either real (for some range of $k_y$) with $\mathcal{PT}$-symmetrical eigenmodes or purely imaginary (in the complementary range of $k_y$), in which case the eigenmodes break the $\mathcal{PT}$ symmetry. The boundaries between the real and imaginary admittances are defined by the EPs, where all the eigenmodes coalesce at the eigenvalue of zero.  Two of the four EPs are located at $k_y=\mp\pi\pm\arccos( R_g/2C_y)$ while the other two are at $k_y=\pm\arccos( R_g/2C_y)$. As the magnitude of $R_g$ increases, the range of $k_y$ corresponding to the real (imaginary) part of the admittance spectrum shrinks (expands). At some critical value given by $R_c=2C_y$, the whole spectrum becomes purely imaginary  with a thre EPs (see Fig. \ref{gFig1}c). When $R_g$ exceeds $R_c$, the two admittance bands will become gapped and no EP exists in the Brillouin zone (not shown in Fig. \ref{gFig1}). In this case, the admittance eigenmodes break the $\mathcal{PT}$ symmetry for the entire range of $k_y$ in the BZ. As can be seen from Fig. \ref{gFig1}c, the real part of the admittance spectrum vanishes at the critical resistance $R_c$. In summary, we can obtain a variable number of EPs depending on the $R_g$ parameter, i.e., two, four, three, and zero EPs for $R_g=0$,  $R_g< 2C_y$, $R_g=2 C_y$, and $R_g> 2C_y$, respectively.
	
\begin{figure}[ht!]
\centering
\includegraphics[scale=0.45]{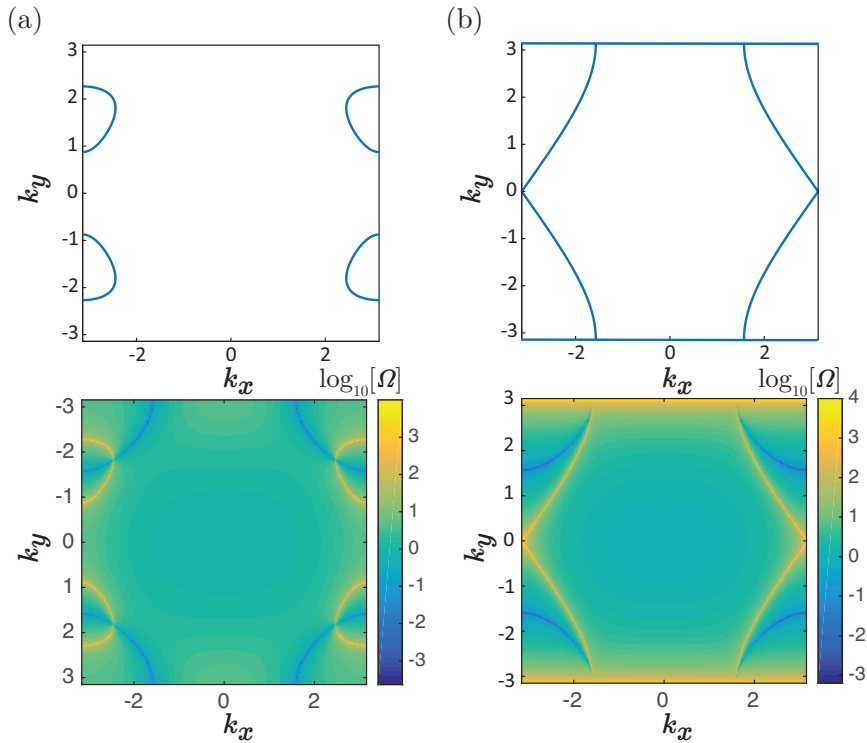}
\caption{The top panels show the loci of the exceptional points in 3D non-Hermitian systems, i.e., systems with finite resistive couplings, at $C_1=0.78$ $\mathrm{mF}$, $C_y=0.39$ $\mathrm{mF}$, $k_z=\pi/2$, and a. $R_g=0.5~\mathrm{mF}$ , and b. $R_g=0.78\ \mathrm{mF}$. The lower panels show the corresponding impedance spectra for the corresponding three-dimensional systems with the same values of $C_1, C_y$ and $R_g$ as the two-dimensional systems. (Note that the quantities plotted are the base 10 logarithms of the impedances in Ohms.)  }
\label{gFig2}
\end{figure}

To obtain the exceptional lines (the loci of the exceptional points), we use the equation for the degeneracy points of the admittance spectrum:
\begin{equation}
   \left(C_1(1+\cos{k_x} )+ 2C_y\cos{k_y}\right)^2 +(C_1\sin{k_x})^2= R_g^2. 
   \label{eq3}
\end{equation}
Eq. \ref{eq3} governs the loci of the exceptional points in the $k_x$-$k_y$ plane. A finite $R_g$ will transform a single pair of band-touching points into exceptional or nodal lines on the $k_z = 0$ plane characterized by Eq. \ref{eq3}. On these exceptional lines, both the real and imaginary parts of the eigenvalues vanish. The top panels of Fig. \ref{gFig2}a and b show the exceptional lines for the parameter sets in Fig. \ref{gFig1}b and c , respectively.  The exceptional points shown in Fig. \ref{gFig1}b and c then correspond to the $k_y$ cross sections of the exceptional lines in Fig. \ref{gFig2} at $k_x=\pi$. 
In addition to the above analysis of the exceptional lines based on the admittance spectrum, an alternative visual representation of the exceptional lines can also be obtained from the impedance spectrum of the TE circuit. In general, the impedance between any two arbitrary nodes $p$ and $q$ in the circuit can be measured by connecting an external current source providing a fixed current  $I_{pq}$ to the two nodes and measuring the resulting voltages at the two nodes $V_p$ and $V_q$. The impedance in the circuit is then given by
\begin{equation}
Z_{pq}=\frac{V_p-V_q}{I_{pq}}=\sum_i \frac{|\psi_{i,p}-\psi_{i,q}|^2}{\lambda_i},
\label{eq4}
\end{equation}
where  $\psi_{i,a}$ and $\lambda_i$  are the voltage at node $a$ and the eigenvalue of  the $i^{\mathrm{th}}$ eigenmode of the (finite-width) Laplacian. One of the key characteristics of Eq. \ref{eq4} is that the impedance diverges (increases to a large value) in the vicinity of the zero-admittance modes ($\lambda_i=0$ ) for non-zero eigen-mode voltages $\psi_{i,p}$ and $\psi_{i,q}$. Therefore, the locus of the high impedance readout can be used to mark out the exceptional lines in momentum space. The lower panels of Fig. \ref{gFig2}a and b depict the corresponding impedance spectra for the parameter sets in Fig. \ref{gFig1}b and c respectively. The plotted impedance is that across the two nodes in a unit cell (i.e. with nodes $p$ and $q$ chosen to be terminal points at either end of the circuit). (For the case of $C_1 = 2 C_y = R_g$ plotted in Fig. \ref{gFig2}b, the $k_y=\pm \pi$ lines are also exceptional lines.)  For both resistive values, the locus of high impedance readouts coincides exactly with the exceptional lines in the admittance spectrum (compare upper and lower panels of Fig. \ref{gFig2}). This suggests the possible electrical detection of exceptional lines in the TE system via impedance measurements.
\subsection{Zero-admittance states in finite system}
\begin{figure}[ht!]
\centering
\includegraphics[scale=0.45]{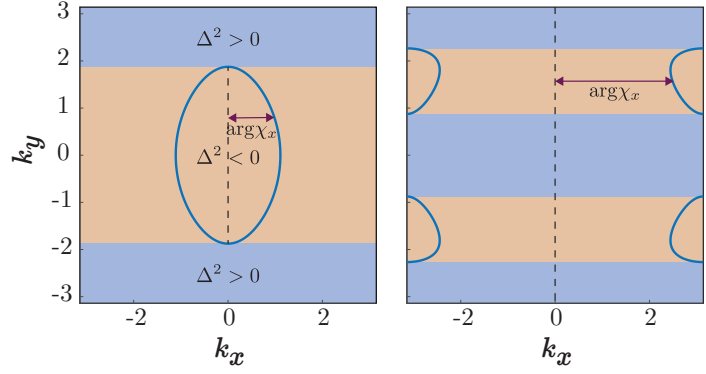}
\caption{ The equal admittance contours (EACs) at $E=0$ for (left) with  parameters $C_1=0.78$ $\mathrm{mF}$, $C_y=0.1$ $\mathrm{mF}$, and $R_g=1.5 $ $\mathrm{m\Omega^{-1}{Hz}^{-1}}$ , and (right) with parameters $C_1=0.39$ $\mathrm{mF}$, $C_y=0.78$ $\mathrm{mF}$, $R_g= 0.0 $ $\mathrm{m\Omega^{-1}{Hz}^{-1}}$ . The regions in the Brillouin zone where $\Delta^2 > 0$ and $\Delta^2 < 0$ are indicated.} 
\label{gChiX1}
\end{figure}

\begin{figure}[ht!]
\centering
\includegraphics[width=0.78\textwidth]{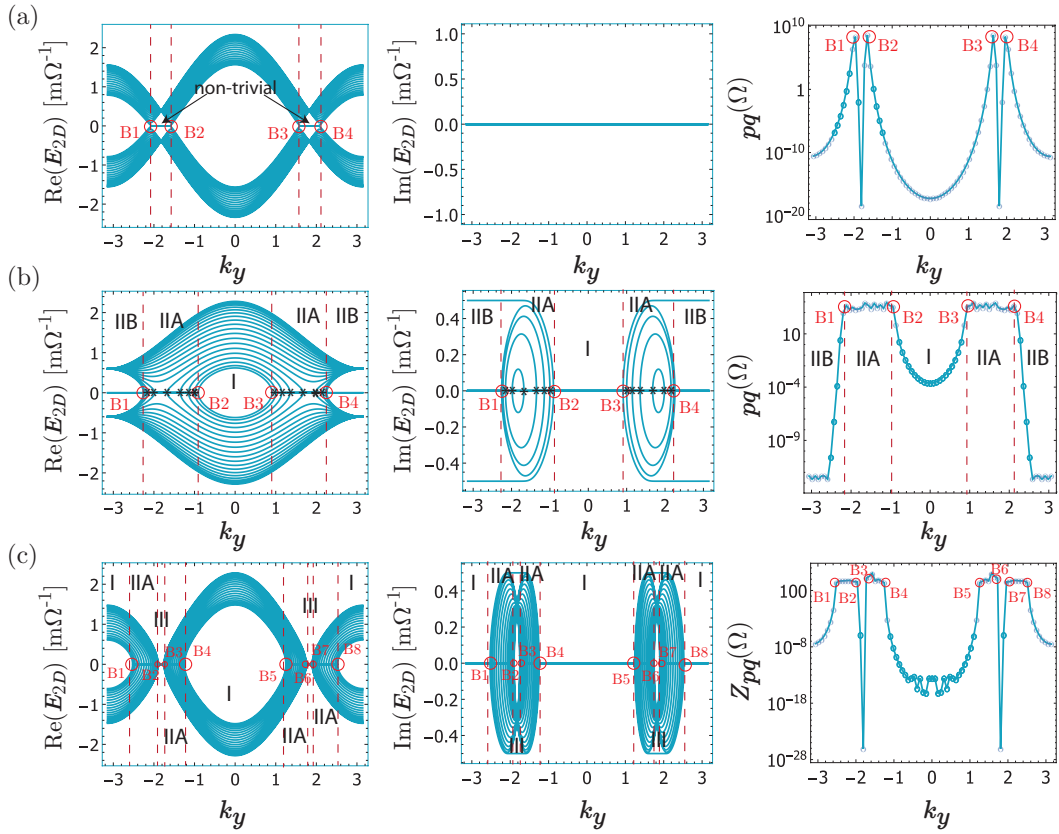}
\caption{Evolution of edge states and boundary points (BP) which mark the transition between non-zero and zero admittance surface states in a non-Hermitian TE circuit, as the resistive and capacitive couplings are varied. We consider a finite TE circuit with $N=\mathrm{20}$ unit cells along the $x$-direction. The first and second columns depict the real and imaginary parts of the admittance spectra, while the third column plots the corresponding spectra for the impedance taken between two terminal points. Parameters used for Panel a: $C_1=0.39$ $ \mathrm{mF}$, $C_y=0.78$ $ \mathrm{mF}$, and $R_g=0 $ $\mathrm{mF}$; Panel b: $C_1=0.78$ $ \mathrm{mF}$, $C_y=0.39$ $ \mathrm{mF}$, and $R_g=0.5 $ $\mathrm{mF}$; and Panel c: $C_1=0.39$ $ \mathrm{mF}$, $C_y=0.78$ $ \mathrm{mF}$, and $R_g=0.5 $ $\mathrm{mF}$. All the BPs are marked with open red circles in the admittance dispersion and impedance spectra. The $k_y$ regions between the BPs are denoted as  I, III, IIA and IIB regions for ease of reference in the text. } 
\label{gFig4}
\end{figure}

To gain further insight into the zero-admittance states of a non-Hermitian system, we will study a 2D dissipative TE system described by Eq. \ref{WSMham} , which is finite along the $x$-direction, i.e., having open boundary conditions along that direction, but is infinite in the $y$-direction. Before investigating the properties of the finite system, we will first explain some properties of the \textit{infinite}-sized 2D system. For this system, Kirchoff's current law at the $A$ and $B$ nodes at resonance can be written as
\begin{equation}
-E V_{x,y}^A = -C_1 V_{x,y}^B + C_1 V_{x-1,y}^B - C_y (V_{x,y+1}^B + V_{x,y-1}^B)+i R_g V_{x,y}^A,
\label{eq8}
\end{equation}
and
\begin{equation}
-E V_{x,y}^B = -C_1 V_{x,y}^A + C_1 V_{x+1,y}^A - C_y (V_{x,y+1}^A + V_{x,y-1}^A)-i R_g V_{x,y}^B.
\label{eq9}
\end{equation}
By substituting the ansatz $ V_{x,y} = \lambda e^{i k_x + i k_y}$ in Eqs. \ref{eq8} and \ref{eq9}, we obtain
\begin{equation}
(E-i R_g\sigma_z)|\lambda\rangle  = (C_1 (1 + \chi_x (\sigma_x-i \sigma_y)+\chi_x^{-1} (\sigma_x+i\sigma_y)) + 2 C_2 \cos{k_y}\sigma_x)|\lambda\rangle 
\label{eq10}
\end{equation}
where $\chi_x= e^{i k_x}$ and $\lambda=(\lambda_A, \lambda_B)^{\mathrm{T}}$. For a given $E$ and $k_y$, $\chi_x$ can be solved from Eq. \ref{eq10} as 
\begin{equation}
\chi_x =\frac{-(t^2+C_1^2-p^2)\pm \sqrt{\Delta^2 }}{2 t C_1},
\label{eq12}
\end{equation}
where $t=C_1+2C_y \cos{k_y}$ and $p^2=E^2+R_g^2$,  and 
\begin{equation}
\Delta^2\equiv(t^2+C_1^2-p^2)^2 - (2tC_1)^2.  \label{Delta2}
\end{equation}
When $\Delta^2 < 0$, 
\begin{equation}
	\chi_{x;(\Delta^2<0)} =\frac{-(t^2 + C_1^2 - p^2) \pm i \sqrt{  |\Delta^2|}}{2tC_1}. \label{chiReal}
\end{equation}
It can be readily seen that $|\chi_{x;(\Delta^2<0)}|=1$, and because $\chi_x \equiv \exp(i k_x)$, this indicates that the corresponding values of $k_x$ would be real when $\Delta^2 < 0$. In this case, $k_x = \arg(\chi_{x;(\Delta^2 < 0)}) = \pm \arctan( \sqrt{(-\Delta^2)}/ (t^2+C_1^2-p^2))$. On the other hand, when $\Delta^2 >0$, $\chi_{x;(\Delta^2 > 0)}$ is real, and, in general,  $|\chi_{x;(\Delta^2 > 0)}| \neq 1$. This indicates that the corresponding values of $k_x$ would be imaginary. At the boundary between the two cases, we have
\begin{eqnarray*}
	\Delta^2 = 0 &\Rightarrow& (t^2+C_1^2-p^2)^2 = (2tC_1)^2 \\
	&\Rightarrow& \chi_{x;(\Delta^2=0)} = -\frac{\sqrt{(2t C_1)^2}}{2t C_1} = -\mathrm{sign}(2t C_1),
\label{deltazero}
\end{eqnarray*}
so that the corresponding $k_x = 0, \pi$ depending on the sign of $2t C_1$. 
When $\Delta^2 < 0$, the $\pm i\sqrt{|\Delta^2|}$ terms in Eq. \ref{chiReal}  result in a finite separation along the $k_x$ axis between two points on the same equal admittance contours (EACs)  for a given $k_y$ value (see Fig. \ref{gChiX1}). The two values of $k_x$ meet when $\Delta^2 = 0$. Figure \ref{gChiX1}a depicts the case where $\chi_x = +1$ when $\Delta^2=0$. Here, for a given $k_y$, $k_x$ on the EAC becomes single-valued at $k_x=0$. Figure \ref{gChiX1}b depicts the case where $\chi_x = -1$ when $\Delta^2=0$, and here $k_x$ becomes single-valued at $k_x=\pi$. For both cases, the values of $k_y$ where $\Delta^2=0$ mark the boundaries for the existence of real $k_x$.
We now consider the nanoribbon geometry with a finite width along the $x$-direction. We show in the Supplementary Materials that for the nanoribbon geometry with non-zero $|R_g|$, the zero-admittance exceptional points do not occur within the bulk energy gaps, but in the bulk bands where real values of $k_x$ exist for $E=0$ in the infinite bulk system. Due to the finite width of the nanoribbons, these zero-admittance points occur as quantized bulk states. The points on the $k_y$ axis that mark the threshold for the existence of the zero admittance states for the nanoribbon system coincide with the $k_y$ values where the solutions of $\chi_x$ at $E=0$ in Eq. \ref{eq12} are equal, i.e., when $\Delta^2=0$. These values of $k_y$ would mark the `boundary points' (BPs) of the system. The values of $k_y$ corresponding to the BPs are given by the solutions of the following equation:
\begin{equation}
C_1+2 C_y \cos{k_y}+\eta C_1=\zeta R_g,
\label{eq13}
\end{equation}
where $\eta=\pm 1$ and $\zeta=\pm 1$ independently. The four possible combinations of $\eta$ and $\zeta$ and the two possible signs of $k_y$ in Eq. \eqref{eq13} therefore provide up to eight real solutions for $k_y$. The BPs come in pairs, and hence, depending on the choice of the coupling capacitances $C_1$ and $C_y$ and resistance $R_g$, we can obtain up to a maximum of four pairs of BPs (see later discussions on phase diagram). Note that the BPs set the boundaries for the existence of exceptional points. In a nanoribbon, the exceptional points do not necessarily appear exactly at the BPs, but in between alternating pairs of BPs.  To illustrate the role of the capacitive and resistive coupling parameters in determining the number of BPs, we plot the admittance spectra for finite TE circuits with $N=20$ unit cells along the $x$-direction in Fig. \ref{gFig4}. 
In the Hermitian limit (i.e. $R_g=0$), the admittance spectrum is purely real, and each pair of quantized bands is symmetric about the $E=0$ axis. The Hermitian TE circuit can host two or four BPs depending upon the relative strength of the capacitive couplings. If $C_1> C_y$, we only have a pair of BPs occurring at $k_y=\pm\pi/2$. However, an additional pair of BPs emerges at $k_y=\arccos(\pm C_1/C_y)$ in the spectrum if $C_1<C_y$ (which is the case illustrated in Fig. \ref{gFig4}a). The Hermitian system possesses chiral and time-reversal symmetries and belongs to to the BDI Altland-Zimbauer class \cite{kawabata2019symmetry}. Treating $k_y$ as a parameter to an effectively one-dimensional model, the effective 1D model along the $x$ direction is mathematically identical to the SSH model, for which the topological invariant is the winding number. The band-touching points B1 to B4 will then correspond to the values of $k_y$ at which the bulk band gap closes and the system transits between topologically trivial phases and topologically non-trivial phases. The emergence of the non-trivial states in Fig.  \ref{gFig4}a) span over ($-\pi/2$ to $ -\arccos( C_1/C_y) $) and ($\pi/2$ to $ \arccos( C_1/C_y) $)  in the $k_y$ axis. (Although the bands appear to be flat at the scale of the figure, they actually disperse weakly and touch only at isolated points.)  The transition points between non-trivial (with edge states) and trivial regions (without edge states) are marked by the onset of high impedance states (shown in the rightmost plot of Fig. \ref{gFig4}a).
A finite non-Hermitian gain or loss term $R_g$ results in four different types of admittance regions for a given value of $k_y$. These admittance regions can host purely real, purely imaginary, complex spectra with exceptional points, and complex admittances without $ E_{2D}=0$ states, which are labelled as  I, III, IIA and IIB  respectively, as shown in Figs. \ref{gFig4}b and \ref{gFig4}c. (For instance, the states between B2 and B3, and B6 and B7 in Fig. \ref{gFig4}c have purely imaginary admittances.)   Eigenstates with preserved and broken $\mathit{PT}$ symmetry are present in regions I and III. Additionally, region IIA hosts exceptional points while IIB does not (see Figs. \ref{gFig4}b and \ref{gFig4}c ). The difference between regions IIA and IIB can be explained via the existence and non-existence of real solutions of $\chi_x$ in Eq. \ref{eq12}. The transition points between any two successive regions are marked by the BPs. Moreover, the zero-admittance states in region IIA  and the $E_{2D}\neq0$ states in the other regions are characterized by high and low impedances respectively (see impedance plots shown in the right-most column of Fig. \ref{gFig4}b and c ). For illustration, the EPs in the region IIA are marked by crosses in Fig. \ref{gFig4}b in both the impedance and admittance plots.  The impedance readout switches between the high and low impedance states occur at the BPs, which mark the boundaries between the IIA and other regions. 
\begin{figure}[ht!]
\centering
\includegraphics[width=0.70\textwidth]{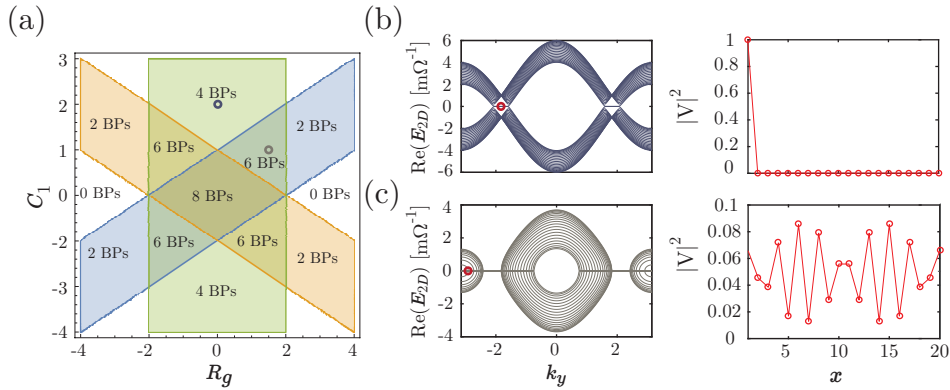}
\caption{a) Phase diagram of a 2D non-Hermitian TE model showing the effect of the capacitive coupling $C_1$ and the gain/loss parameter $R_g$ on the number of BPs, and hence the number of zero-admittance regions in the admittance spectrum. We consider a finite number of unit cells along $x$-direction, and set $C_y=1.0$ $ \mathrm{mF}$. b) and c): (Left) Dispersion relations of $N_x=20$ unit cell-wide nanoribbons at b. $C_1=2$ mF and $R_g = 0$ mF and c. $C_1=1\ \mathrm{mF}$ and $R_g=1.5\ \mathrm{mF}$ indicated by the two circles in panel a; (Right) The square of the voltage amplitudes corresponding to the EP with the minimum $k_y$ value, as indicated by the circles in the dispersion relations.}  
\label{gFig6}
\end{figure} 
Fig. \ref{gFig6}a shows the phase diagram of the 2D non-Hermitian TE model as a function of the resistive coupling $R_g$ and the capacitive coupling $C_1$. The different phases are characterized by different numbers of BPs and hence different numbers of zero-admittance regions. These phases are separated from one another by phase boundaries that are defined by the existence and number of real solutions in Eq. \ref{eq13}. Note that BPs always occur and annihilate in pairs and they connect two admittance bands together. 
To study the effect of non-Hermitian term $R_g$ on the zero-admittance states, we consider a 2D TE system with open-boundary conditions in the $x$-direction. In the Hermitian condition, i.e., $R_g=0$, the system hosts zero-energy modes in which the square of the voltage amplitudes are localized at its edges, as shown in top panel of Fig. \ref{gFig6}b. This corresponds to the usual non-trivial SSH edge state. However, in the non-Hermitian case, i.e., with the introduction of a finite $R_g$, we would instead obtain bulk states at $E=0$ rather than edge states (see Fig. \ref{gFig6}c). This is indicated by the square of the voltage amplitude being no longer localized at an edge.  
\subsection{Conclusion}
In conclusion, we proposed a topolectrical (TE) circuit model with resistive elements which provide loss and gain factors that break the Hermiticity of the circuit to model and realize various non-Hermitian topological phases. By varying the resistive elements, the loci of the exceptional points (or exceptional lines) of the circuit can be modulated. IWe showed that the topology of the exceptional lines in the Brillouin zone can be traced by the impedance spectra of the circuit. Additionally, we studied a finite TE system where open boundary conditions apply in one of the dimensions. In these finite circuits, we demonstrated the tunability of both the number of exceptional points corresponding to zero-admittance states, as well as that of boundary points (BPs) which delineate the circuit parameter range where these exceptional points exist.  The regions separated by the BPs are characterized by high and low values of impedance differing by several orders of magnitude, which are detectable in a practical circuit. We also derived a phase diagram of the finite TE system which delineates different between topological phases that are characterized by different number of BP pairs (up to a maximum of four). The edge state character of zero-admittance states of Hermitian LC circuits are transformed into exceptional points which are hybridized with the bulk modes in the non-Hermitian RLC circuits. In summary, we have proposed a tunable electrical framework consisting of RLC circuit networks as a means to realize different topological phases of non-Hermitian systems, and characterize them based on their impedance output, as well as their BP and exceptional line configurations. 
\subsection*{Data availability}
The data that support the plots within this paper and other
findings of this study are available from the corresponding author upon reasonable
request.
\subsection*{Code availability}
The computer codes used in the current study are accessible from the corresponding author upon reasonable request.
\subsection*{Acknowledgement}
This work is supported by the Ministry of Education (MOE) Tier-II grant MOE2018-T2-2-117 (NUS Grant Nos. R-263-000-E45-112/R-398-000-092-112), MOE Tier-I FRC grant (NUS Grant No. R-263-000-D66-114), and other MOE grants (NUS Grant Nos. C-261-000-207-532, and C-261-000-777-532).

\subsection*{Author contributions}
S.M.R-U-I, Z.B.S and M.B.A.J initiated the primary idea. S.M.R-U-I and Z.B.S
contributed to formulating the analytical model and developing the code, under the kind supervision of M.B.A.J. All the authors contributed to the data analysis and the writing of the manuscript.  
\subsection*{Additional information}
\textbf{Competing interests:}  The authors declare no competing interests.
%

\end{document}